\documentclass{article}
\usepackage{ijcai23}

\usepackage{times}
\usepackage{soul}
\usepackage{url}
\usepackage[hidelinks]{hyperref}
\usepackage[utf8]{inputenc}
\usepackage[small]{caption}
\usepackage{graphicx}
\usepackage{amsmath}
\usepackage{amsthm}
\usepackage{booktabs}
\usepackage{algorithm}
\usepackage{algorithmic}
\usepackage[switch]{lineno}

\usepackage{subfigure}
\usepackage{amssymb}
\title{Sync Pure Counterfactual Regret Minimization in Incomplete Information Extensive Form Games}

\author{
    Ju Qi
    \affiliations
    Huazhong University of Science and Technology
    \emails
    juqi@hust.edu.cn
}

\begin{document}

\maketitle

\begin{abstract}
    Counterfactual Regret Minimization (CFR) and its variants developed based upon Regret Matching (RM) have been considered to be the best method to solve incomplete information extensive form games.
    In addition to RM and CFR,
    Fictitious Play (FP) is another equilibrium computation algorithm in normal form games.
    Previous experience has shown that the convergence rate of FP is slower than RM and FP is difficult to use in extensive form games.
    However, recent research has made improvements in both issues.
    Firstly, Abernethy proposed a new FP variant sync FP, which has faster convergence rate than RM+.
    Secondly, Qi introduced FP into extensive form games and proposed Pure CFR (PCFR).
    This paper combines these two improvements,
    resulting in a new algorithm sync PCFR.
    In our experiment,
    the convergence rate of sync PCFR is approximately an order of magnitude faster than CFR+ (state-of-the-art algorithm for equilibrium computation in incomplete information extensive form games),
    while requiring less memory in an iteration.
\end{abstract}

\section{Introduction}\label{sec:introduction}
    %! Author = juqi
%! Date = 2023/11/8

Currently,
almost all achievements in two-player zero-sum extensive form games come from Counterfactual Regret Minimization (CFR)~\cite{2007Regret} and its variants including~\cite{2010Monte},\cite{2014Solving},\cite{2018Lazy},\cite{2019Solving}.
Fictitious Play (FP)~\cite{2007Brown},\cite{1951Iterative} is another no regret learning algorithm,
but it is inferior to CFR in terms of application range and convergence rate.
Specifically:
\begin{itemize}
    \item The convergence rate of vanilla FP is not as good as RM~\cite{brown2019deep}, and there are some \textit{special cases}~\cite{daskalakis2014counter} that cause the convergence rate of FP to drop significantly.
    \item Most of the research on FP is conducted in normal form games. Compared with CFR, the previous FP algorithm in extensive form games will greatly increase the time and space complexity~\cite{1996Fictitious},\cite{2015Fictitious}.
\end{itemize}

But recent research has made improvements on both issues.
For slow convergence,
Abernethy~\cite{abernethy2021fast} proposed sync FP.
This method not only overcomes the slow convergence rate in \textit{special case} by lexicographic order,
and it is proved that sync FP can converge to the equilibrium at a rate of $O\left(\frac{L}{\sqrt{T}}\right)$ in the diagonal payoff matrices.
For high complexity,
Qi~\cite{qi2023pure} introduced FP into the CFR process and obtained the Pure CFR (PCFR) algorithm,
which proved that replacing the regret matching strategy in CFR with the best responce strategy will not affect the convergence property.
Compared with CFR,
PCFR can reduce memory requirements and has similar convergence rate.

In this technical report,
we combine sync FP and PCFR and propose a new algorithm called sync PCFR.
This algorithm can perfectly combine the advantages of PCFR and sync FP,
and only requires few changes on the basis of CFR.
Experiments show that the convergence rate of sync PCFR is approximately an order of magnitude faster than CFR+,
the best variant of CFR.
At the same time,
compared with CFR+,
the memory requirements of sync PCFR in one iteration is greatly reduced.

\section{Preliminaries}\label{sec:preliminaries}
    %! Author = juqi
%! Date = 2023/11/8

\subsection{Vanilla Fictitious Play}\label{subsec:vanilla-fictitious-play}
Define $b^i(\sigma^{-i})=\arg\max_{a^i\in A^i}u^i(a^i,\sigma^{-i})$ as the pure best response (BR) strategy of player $i$ to remaining players’ strategy profile $\sigma^{-i}$,
and $b(\sigma)=\times^N_{i=1}b^i(\sigma^{-i})$ is pure BR strategy profile to the mixed strategy profile $\sigma$.
In the vanilla FP process,
assuming that all players start from the random strategy profile $\bar\sigma_{t=1}$,
then the average strategy profile update follows:
\begin{equation}
    \bar\sigma_{t+1}=\left( 1-\frac{1}{t+1}\right)\bar\sigma_t+\frac{1}{t+1}b(\bar\sigma_t),
    \label{eq:201}
\end{equation}
where $t$ represents the number of iterations.
$\bar\sigma_t$ converges to Nash Equilibrium (NE) with $t\rightarrow\infty$.
The convergence rate of vanilla FP is $O\left(\frac{L}{\sqrt[2|A|-2]{T}} \right)$~\cite{robinson1951iterative},
which is much slower than RM $O\left(\frac{L\sqrt{|A|}}{\sqrt{T}} \right)$.

\subsection{Sync Fictitious Play}\label{subsec:sync-fictitious-play}
Karlin~\cite{karlin1959mathematical} suggested that the rate of FP may be on the order of $O\left( \frac{L}{\sqrt{T}}\right)$.
Recently Abernethy~\cite{abernethy2021fast} proposed a new method sync FP,
which proved Karlin's conjecture in diagonal payoff matrices.
The motivation of sync FP is:
During the FP process,
the BR strategy does not change in most iterations,
this unchanged BR strategy profile is called sync phase,
and convergence can be accelerated by skipping the sync phase.

\begin{table*}[h]
    \centering
    \includegraphics[width=0.9\textwidth]{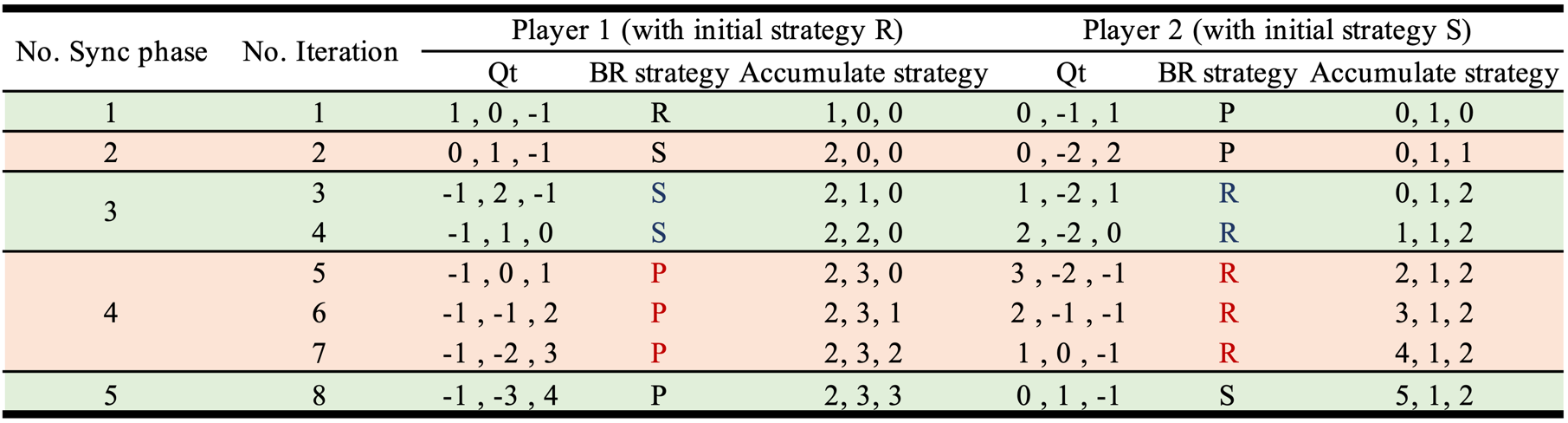}
    \caption{
        Sync phase in rock-paper-scissors.
    }
    \label{table:syncphase}
\end{table*}

As shown in the Table~\ref{table:syncphase},
the BR strategy profiles at $t=2\sim 3$ and $t=4\sim 7$ have not changed.
Abernethy proved that sync FP can achieve a convergence rate of $O\left( \frac{L}{\sqrt{T}}\right)$ in diagonal payoff matrices,
and seems to have similar convergence rates in other forms of payoff matrices.

\subsection{Pure CFR}\label{subsec:pure-cfr}
Qi~\cite{qi2023pure} introduced FP on the basis of CFR and obtained PCFR.
Firstly,
there is no need to calculate the regret value in PCFR.
Define immediate Q-value as
\begin{equation}
    Q_{t}^{i}(I, a)=Q_{t-1}^{i}(I, a)+\pi_{\sigma_{t}}^{-i}(I)u^{i}\left( I, \sigma_{t}|_{I \rightarrow a}\right).
    \label{eq:202}
\end{equation}

The difference between $Q(I)$ and regret ${R}(I)$ is that there is no need to subtract the average strategy payoff $u^ {i}\left(I,\sigma_{t} \right)$.
Secondly,
the strategy in the next iteration is a pure BR strategy rather than mixed regret matching strategy:

\begin{equation}
    \sigma^i_{t+1}={\arg\max}_{a\in A}Q_{t}^{i}(I, a).
    \label{eq:203}
\end{equation}

In this case,
not only does the step of solving the next iteration strategy become simpler,
but also naive pruning can be triggered with the highest efficiency.
This allows PCFR to pass through very few nodes in an iteration, thereby saving memory.

\section{Algorithm}\label{sec:algorithm}
    %! Author = juqi
%! Date = 2023/11/8

\subsection{Q-value Based Sync Fictitious Play}\label{subsec:q-value-based-sync-fictitious-play}
Based on sync Fictitious Play we propose a new sync FP implementation Q-value based sync FP.
We regard solving sync phase length as a pursuit problem.
Define $Q_t^i(a)$ as the Q-value of player $i$ at time $t$,
and let $a^*={\arg\max}_{a\in A}Q^i_t(a)=b^i(\sigma^{-i})$.
The distance that other strategies need to pursue is
\begin{equation}
    Q_t^{\text{gap},i}(a)=Q^i_t(a^*)-Q^i_t(a).
    \label{eq:301}
\end{equation}
the pursuit speed per iteration is
\begin{equation}
   S^i(a)=u^{i}\left( a,\sigma^{-i}\right)-u^i\left(a^*, \sigma^{-i}\right),
    \label{eq:3011}
\end{equation}
if $S^i(a)>0$ and $Q_t^{\text{gap},i}(a)>0$,
the pursue time $w^i_{\text{pst}}$ of strategy $a$ can be calculated
\begin{equation}
    w^i_{\text{pst}}(a)=
    \begin{cases}
        \left\lceil \frac{Q_t^{\text{gap},i}(a)}{S^i(a)}\right\rceil & \text { if } S^i(a)>0, Q_t^{\text{gap},i}(a)>0 \\
        \inf & \text { otherwise,}
    \end{cases}
    \label{eq:302}
\end{equation}
where $\lceil\cdot\rceil$ represents rounding up.
At this time, the sync phase length of player $i$ is

\begin{equation}
    w^i_{\text{pst}}=\min_{a\in A}w^i_{\text{pst}}(a).
    \label{eq:303}
\end{equation}

We choose a smaller length of sync phase between player 1 and player 2 $w_{\text{pst}}=\min_{i\in \{1,2\}}(w^i_{\text{pst}})$,
then skip the sync phase
\begin{equation}
    \bar\sigma_{T+1}=(1-\frac{1}{T+w_{\text{pst}}})\bar\sigma_T+\frac{1}{T+w_{\text{pst}}}b(\bar\sigma_t),
    \label{eq:304}
\end{equation}

\begin{equation}
    Q_{T+1}(a)=Q_{T}(a)+w_{\text{pst}}u^i(a,\sigma^{-i}_t).
    \label{eq:305}
\end{equation}

\subsection{Sync PCFR}\label{subsec:sync-pcfr}
Since the essence of PCFR is a method of using FP in extensive form games, sync phase also exists in PCFR.
The Q-value of player $i$ at time $t$ in information set $I$ is $Q_t^i(I,a)$,
let $a^*={\arg\max}_{a\in A(I)}Q^i_t(I,a)=b^i(\sigma^{-i})(I)$.
The distance that other strategies need to pursue is
\begin{equation}
    Q_{t}^{\text{gap},i}(I,a)=Q^i_t(I,a^*)-Q^i_{t}(I,a).
    \label{eq:306}
\end{equation}
the pursuit speed per iteration in information set $I$ is:
\begin{equation}
    S^i(I,a)=\pi_{\sigma_{t}}^{-i}(I) \left(u^{i}\left( I,\sigma_{t}|_{I \rightarrow a}\right)-u^{i}\left(I,\sigma_{t}|_{I \rightarrow a^*}\right)\right).
    \label{eq:307}
\end{equation}
when $S^i(I,a)>0$ and $Q_{t}^{\text{gap},i}(I,a)>0$ the pursue time of strategy $a$ in information set $I$ can be calculated
\begin{equation}
    w^i_{\text{pst}}(I,a)=
    \begin{cases}
        \left\lceil \frac{Q_t^{\text{gap},i}(a)}{S^i(I,a)} \right\rceil & \text { if } S(I,a)>0, Q_{t}^{\text{gap},i}(I,a)>0 \\
        \inf & \text {otherwise,}
    \end{cases}
    \label{eq:308}
\end{equation}
the length of sync phase is
\begin{equation}
    w_{\text{pst}}=\min_{i\in\{1,2\}}\min_{I\in \mathcal{I}^i}\min_{a\in A(I)}w^ i_{\text{pst}}(I,a),
    \label{eq:309}
\end{equation}
then we can directly skip the sync phase
\begin{equation}
    Q_{t}^{i}(I, a)=Q_{t-1}^{i}(I, a)+w_{\text{pst}}\pi_{\sigma_{t}}^{-i}(I)u^{i}\left( I,   \sigma_{t}|_{I \rightarrow a}\right),
    \label{eq:310}
\end{equation}

\begin{equation}
    \bar{\sigma}_{T}^i(I,a)=\frac{\sum_{t=1}^{T} w_{\text{pst}} \pi_{\sigma_{t}}^{i} (I)\sigma_{t}^i(I,a)}{\sum_{t=1}^{T}w_{\text{pst}} \pi_{\sigma_{t}}^{i}(I)}.
    \label{eq:311}
\end{equation}

Compared with PCFR,
sync PCFR only skip the sync phases and does not affect the convergence properties at all.
The pseudocode is in Algorithm~\ref{alg:algorithm}.

\section{Experimental results}\label{sec:experimental-results}
    %! Author = juqi
%! Date = 2023/11/8

\subsection{Convergence rate of sync PCFR}\label{subsec:convergence-rate-of-sync-pcfr}
Experiments were conducted on Kuhn poker~\cite{1950Simplified},
Leduc poker~\cite{2002Abstraction} and 5 pot Leduc poker~\cite{brown2020equilibrium}.
These poker games are widely used to test the convergence rate of different algorithms.
The settings of CFR and CFR+ refer to~\cite{brown2020equilibrium},
and the settings of PCFR refer to~\cite{qi2023pure}.

\begin{figure*}[htbp]
    \centering
    \subfigure{
        \centering
        \includegraphics[width=0.9\textwidth]{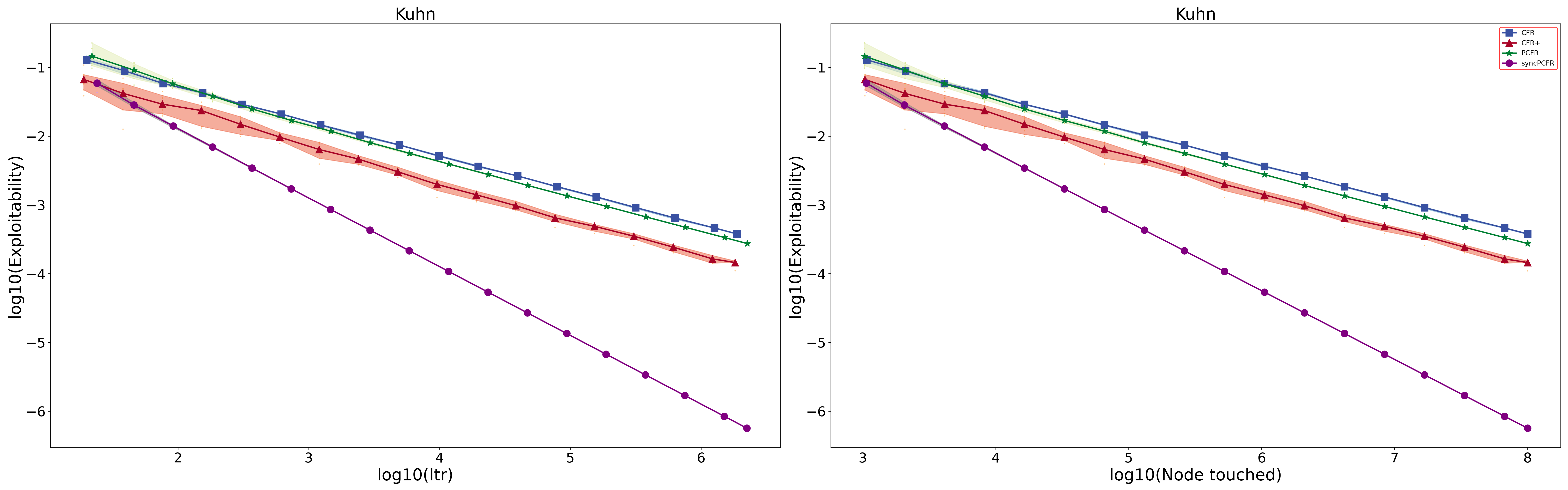}
    }
    \subfigure{
        \centering
        \includegraphics[width=0.9\textwidth]{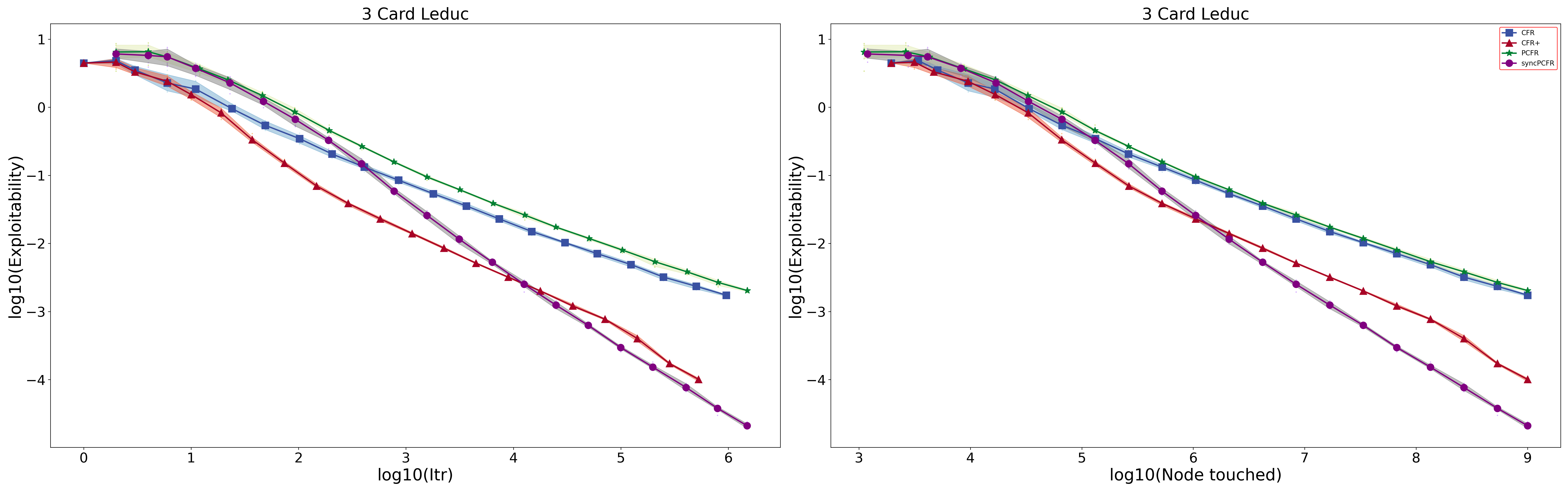}
    }
    \subfigure{
        \centering
        \includegraphics[width=0.9\textwidth]{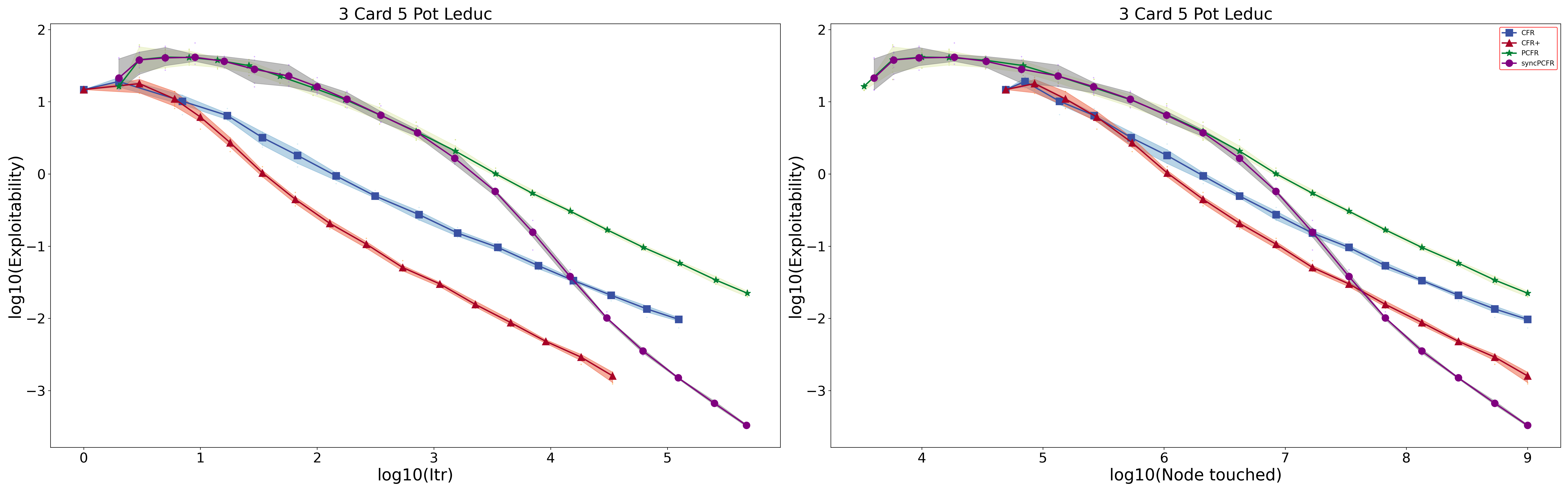}
    }
    \caption{
        Convergence rate in Kuhn, Leduc and 5 Pot Leduc.
        Each experiment has an average of 30 rounds, and the lightrange is the 90\% confidence interval.
    }
    \label{fig:401}
\end{figure*}

It can be seen from the Figure~\ref{fig:401} that in the small-scale game (Kuhn poker),
sync PCFR surpassed the other algorithms in a very short time.
In slightly more complex games,
although the convergence rate of sync PCFR is slower in the early stage of training,
sync PCFR can still surpass CFR+ in the later stages.
More importantly,
since the sync PCFR algorithm can trigger naive pruning with the highest efficiency,
sync PCFR only needs to pass through $\sqrt{|\mathcal{S}|}$ nodes in one iteration,
while CFR/CFR+ passes through all $|\mathcal{S}|$ nodes under the worst case.
Therefore,
when the algorithm performance is measured by the number of nodes touched,
the convergence rate of sync PCFR is approximately an order of magnitude faster than CFR+ (when training time is sufficient).

\subsection{Length of Sync Phase in Extensive Form Games}\label{subsec:length-of-sync-phase-in-extensive-form-game}
Sync PCFR and vanilla PCFR are actually equivalent,
So we need to figure out:
how many calculation steps does sync PCFR skip?

\begin{figure*}[htbp]
    \centering
    \subfigure{
        \centering
        \includegraphics[width=0.9\textwidth]{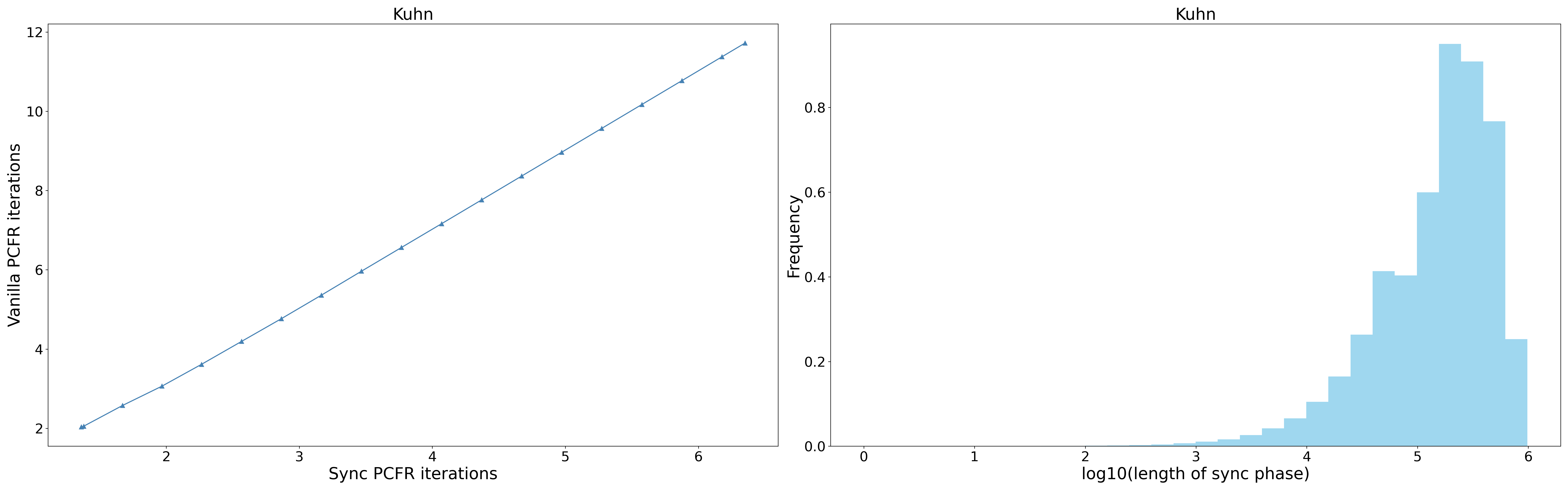}
    }
    \subfigure{
        \centering
        \includegraphics[width=0.9\textwidth]{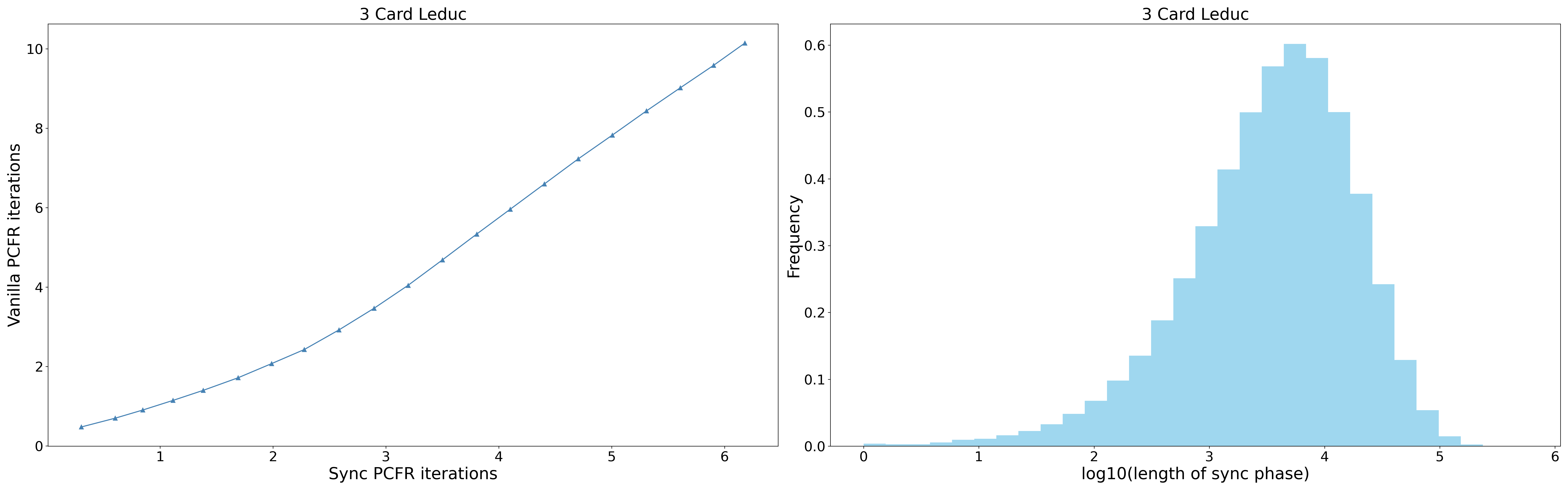}
    }
    \subfigure{
        \centering
        \includegraphics[width=0.9\textwidth]{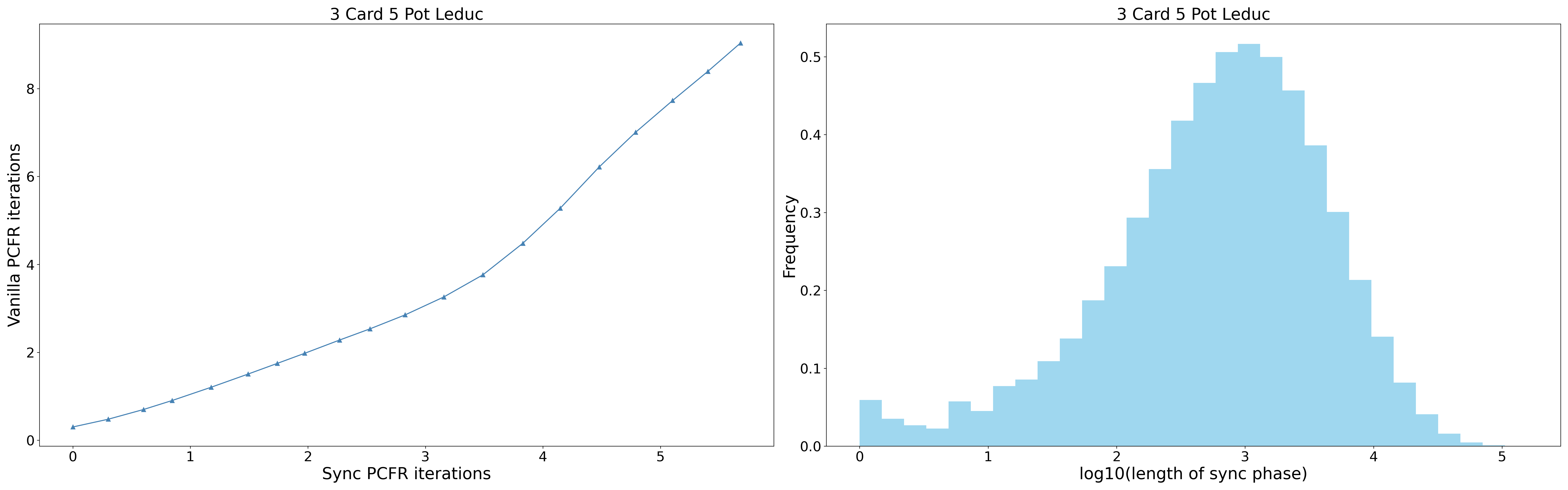}
    }
    \caption{
        The picture on the left is the mapping relationship between sync PCFR and vanilla PCFR itration.
        The picture on the right is the distribution of the length of the sync phase during training.
    }
    \label{fig:402}
\end{figure*}

As shown in Figure~\ref{fig:402},
the number of iterations for sync PCFR is approximately related to the square of the number of iterations for vanilla PCFR,
and the length of the skipped sync phase may follow a log-normal distribution.

\section{Conclusion}\label{sec:conclusion}
    %! Author = juqi
%! Date = 2023/11/8
In this technical reports,
we propose a new algorithm for solving incomplete information extensive form games sync PCFR.
This method has three advantages:
\begin{itemize}
    \item It can be obtained from the previous CFR algorithm with only a few changes;
    \item The convergence rate is greatly improved, approximately an order of magnitude faster than the CFR+;
    \item One sync PCFR iteration requires much less memory than CFR.
\end{itemize}

In future papers we will try to explain the mathematics behind the algorithm in more detail,
as well as explore the use of this method in larger-scale problems.

\appendix

\section*{Acknowledgments}
Thanks to Xuefeng Huang for his suggestions on this technical report.

\bibliographystyle{named}
\bibliography{ijcai23}

\section*{}
    
\begin{algorithm}[t]
    \caption{Sync PCFR}
    \label{alg:algorithm}
    \textbf{Input}: Game and Number of iterations $T$\\
    \textbf{Output}: $\epsilon$-Nash approximation $\bar{\sigma}$
    \begin{algorithmic}[1] %[1] enables line numbers
        \FOR{$I\in \mathcal{I}$}
            \FOR{$a\in A(I)$}
                \STATE $Q(I,a)\leftarrow 0$
                \STATE $\bar\sigma(I,a)\leftarrow 0$\
                \STATE $\sigma(I)\leftarrow$ randomChoice($A(I)$)
            \ENDFOR
        \ENDFOR
        \STATE $w_{\text{pst}}=1$
        \FOR{$t\leftarrow 0$ to $T$}
            \STATE $\sigma\leftarrow$ QValueToStrategy($\mathcal{I},Q$)
            \STATE UpdateAverage($\mathcal{I},\sigma,w_{\text{pst}},\bar\sigma$)
            \STATE $v\leftarrow$ StrategyToValues($\mathcal{I},\sigma$)
            \STATE $w_{\text{pst}}\leftarrow$ UpdateQValues($\mathcal{I},\sigma,v,Q$)
        \ENDFOR
        \STATE NormaliseAverage($\mathcal{I},\bar{\sigma}$)
        \STATE \textbf{return} $\bar{\sigma}$
    \end{algorithmic}

    \textbf{function} QValueToStrategy($UpdateSets,Q$)

    \begin{algorithmic}[1] %[1] enables line numbers
        \FOR{$I \in UpdateSets$}
            \STATE $\sigma(I)\leftarrow {\arg\max}_{a\in A}Q(I, a)$
        \ENDFOR
        \STATE \textbf{return} $\sigma$
    \end{algorithmic}

    \textbf{function} UpdateAverage($UpdateSets,\sigma,w_{\text{pst}},\bar\sigma$)

    \begin{algorithmic}[1] %[1] enables line numbers
        \FOR{$I \in UpdateSets, a\in A(I), i\in \{1,2\}$}
            \STATE $\bar{\sigma}^i(I, a) \leftarrow \bar{\sigma}^i(I, a)+w_{\text{pst}}\pi^i_{\sigma}(I) \sigma^i(I, a)$
        \ENDFOR
    \end{algorithmic}

    \textbf{function} StrategyToValues(\textit{UpdateSets},$\sigma$)

    \begin{algorithmic}[1] %[1] enables line numbers
        \FOR{$I \in UpdateSets,a\in A(I), i\in \{1,2\}$}
            \STATE $v(I, a) \leftarrow \sum_{h \in I \cdot a} \sum_{z \in Z, h \sqsubset z} \pi^{i}_{\sigma}(h) \pi(z \mid h) u^i(z)$
        \ENDFOR
        \STATE \textbf{return} $v$
    \end{algorithmic}

    \textbf{function} UpdateQValues($UpdateSets,\sigma,v,Q$)

    \begin{algorithmic}[1] %[1] enables line numbers
        \FOR{$I \in UpdateSets,a\in A(I)$}
            \STATE $a^*=\sigma(I)$
            \STATE $Q^{\text{gap},i}(I,a)=\max_{a^*\in A(I)}Q^i_t(I,a^*)-Q^i(I,a)$
            \STATE $S^i(I,a)=\pi_{\sigma}^{-i}(I) \left(v(I, a)-v(I, a^*)\right)$
            \IF {$Q^{\text{gap},i}(I,a)>0$,$S^i(I,a)>0$}
                \STATE $w^i_{\text{pst}}(I,a)=\left\lceil\frac{Q_t^{\text{gap},i}(a)}{S^i(I,a)} \right\rceil$
            \ELSE
                \STATE $w^i_{\text{pst}}(I,a)=\inf$
            \ENDIF
        \ENDFOR
        \STATE $w_{\text{pst}}=\min_{i\in\{1,2\}}\min_{I\in \mathcal{I}^i}\min_{a\in A(I)}w^i_{\text{pst}}(I,a)$
        \FOR{$I \in UpdateSets,a\in A(I)$}
            \STATE $Q(I, a) \leftarrow Q(I, a)+w_{\text{pst}}v(I, a)$
        \ENDFOR
        \STATE \textbf{return} $w_{\text{pst}}$
    \end{algorithmic}

    \textbf{function} NormaliseAverage($UpdateSets,\bar\sigma$)

    \begin{algorithmic}[1] %[1] enables line numbers
        \FOR{$I \in UpdateSets$}
            \STATE $sum \leftarrow \sum_{a\in A(I)}\bar\sigma(I,a)$
            \FOR{$a \in A(I)$}
                \STATE $\bar\sigma(I,a) \leftarrow \bar\sigma(I,a)/sum$
            \ENDFOR
        \ENDFOR
    \end{algorithmic}
\end{algorithm}

\end{document}